# Electric field switching of the uniaxial magnetic anisotropy of an antiferromagnet

Xianzhe Chen[1,†], Xiaofeng Zhou[1,†], Ran Cheng[2,†], Cheng Song[1,*], Jia Zhang[3], Yichuan Wu[3], You Ba[4], Haobo Li[4], Yiming Sun[1], Yunfeng You[1], Yonggang Zhao[4], Feng Pan[1,*]

[1]Key Laboratory of Advanced Materials (MOE), School of Materials Science and Engineering, Tsinghua University, Beijing 100084, China

[2]Department of Electrical and Computer Engineering, University of California, Riverside, California 92521, USA

[3]School of Physics and Wuhan National High Magnetic Field Center, Huazhong University of Science and Technology, Wuhan 430074, China

[4]State Key Laboratory of Low-Dimensional Quantum Physics and Department of Physics, Tsinghua University, Beijing 100084, China

Electric field control of magnetic anisotropy in ferromagnets has been intensively pursued in spintronics to achieve efficient memory and computing devices with low energy consumption. Compared with ferromagnets, antiferromagnets hold huge potential in high-density information storage for their ultrafast spin dynamics and vanishingly small stray field. However, the switching of magnetic anisotropy of antiferromagnets via electric field remains elusive. Here we use ferroelastic strain from piezoelectric materials to switch the uniaxial magnetic anisotropy and the Néel order reversibly in antiferromagnetic $Mn_2Au$ films with an electric field of only a few kV/cm at room temperature. Owing to the uniaxial magnetic anisotropy, a ratchet-like switching behavior driven by the Néel spin-orbit torque is observed in the $Mn_2Au$, which can be reversed by electric fields.

---

[†]These authors contributed equally to this work.
[*]E-mail: songcheng@mail.tsinghua.edu.cn, panf@mail.tsinghua.edu.cn

With recent progress in spintronics, magnetization switching driven by spin-transfer torques (*1*) and spin-orbit torques (*2*) has been successfully utilized to fabricate MRAM (*3*) (magnetoresistive random access memory), while electric field (*E*) control of magnetism has been proposed as a building block in MeRAM (*4,5*) (magnetoelectric random access memory) with ultralow energy consumption (*6–8*). Nevertheless, the fundamental operating speed in ferromagnet-based devices is limited to tenth of Gigahertz, and the spatially extended stray field inhibits further increase of memory density. Antiferromagnetic (AFM) materials provide promising alternatives to solve these challenges because of several unique features such as ultrafast spin dynamics and the absence of macroscopic magnetization (*9,10*). To unlock the potential of antiferromagnets, there is a pressing need to control or even switch the AFM moments by electric field. Rotating AFM domains through ferroelectric polarization was observed in multiferroic insulator $BiFeO_3$ (*11*) and $Cr_2O_3$ (*5,12*) through magnetoelectric coupling. Recent developments concerning antiferromagnets have enabled *E*-control of magnetic phase transition and the exchange coupling between antiferromagnets and ferromagnets (*13–15*). However, *E*-switching of the magnetic anisotropy of antiferromagnets is still lacking, where the modulation of magnetic anisotropy can be used to rotate Néel order, regardless of the conducting states. Here we employ ferroelastic strain in ferroelectric materials to accomplish the switching of the uniaxial magnetic anisotropy (UMA) of an antiferromagnet.

Figure 1 illustrates the schematic of nonvolatile ferroelastic switching of Néel moments in antiferromagnetic/ferroelectric heterostructures. We first show in the inset a cartoon for typical *E*-dependent out-of-plane ferroelectric polarization (*P–E*) of Pb(Mg$_{1/3}$Nb$_{2/3}$)$_{0.7}$Ti$_{0.3}$O$_3$ (PMN-PT)(011) crystals. At first the ferroelectric polarization is set to be aligned in-plane after several circulations of *P–E* loop. After applying a positive $E_1$ (larger than the ferroelectric coercive field) on PMN-PT (011), as displayed in Fig. 1A, remanent *P*-vectors would be aligned from in-plane to out-of-plane (*17*). The adjacent AFM films would suffer a compressive strain in the [0$\bar{1}$1] axis and a tensile strain along [100] of the PMN-PT at zero field (fig. S1), with respect to the case with in-plane *P*-vectors. The nonvolatile in-plane strain variation along [0$\bar{1}$1] of PMN-PT(011) is approximately 0.2% (fig. S2). Since the Néel moments intend to be aligned along the compressive axis according to our calculation (figs. S3 and S4) and previous results (*16*), they would be rotated from [100] to [0$\bar{1}$1]. In contrast, when *E* has the opposite polarity up to the ferroelectric coercive field $E_2$ of PMN-PT(011), the remanent *P*-vectors would be rotated from out-of-plane to in-plane (*17*), and the strain state returns to the initial state, accompanied by the Néel moments switching from [0$\bar{1}$1] back to [100] (Fig. 1B). The experiments below demonstrate this concept in AFM Mn$_2$Au, which exhibits a high Néel temperature above 1000 K (*18*), huge AFM anisotropic magnetoresistance, high conductivity for device integration (*19*), a small in-plane anisotropy (~7 μeV) with comparatively low energy barrier for the switching (figs. S3, S4), and Néel spin-orbit torque (NSOT) (*19–21*).

Mn *L*-edge x-ray magnetic linear dichroism (XMLD) spectra were used to detect the Néel order distributions of Mn$_2$Au. Figure 2 presents the corresponding data for 20 nm-thick Mn$_2$Au (103) films grown on 0.5 mm-thick PMN-PT (011) substrate (fig. S5). These spectra were recorded *in-situ* at zero-field after applying the electric field marked in each panel of Fig. 2A. The x-ray was vertically incident to the film and the polarized direction of the x-ray is parallel to PMN-PT(011). XMLD signals are obtained as $\text{XMLD} = \text{XAS}_{//} - \text{XAS}_{\perp}$, where $\text{XAS}_{//}$ and $\text{XAS}_{\perp}$ denote the x-ray absorption spectroscopy (XAS) recorded with [100] (//) and [0$\bar{1}$1] ($\perp$) polarization, respectively (fig. S6). For the as-grown $Mn_2Au$, the Mn $L_3$-edge XMLD spectrum exhibits a zero-negative-positive-zero feature. This is quite characteristic for the Néel order along the parallel (//) direction (*22*), indicating that the $Mn_2Au$ exhibits UMA with the easy-axis along [100] for the original state, which can be ascribed to the asymmetry between [100] and [0$\bar{1}$1] of PMN-PT(011). The scenario differs dramatically after gating $E = +4$ kV/cm. After removing $E$, the Mn $L$-edge XMLD spectrum was recorded, which shows an opposite polarity at $L_3$, e.g., zero-positive-negative-zero, suggesting that the UMA switching with the Néel order rotation from [100] to [0$\bar{1}$1]. Identical measurements were carried out after applying $E = -2$ kV/cm. The Mn $L_3$-edge again undergoes a zero-negative-positive-zero signal, showing the main features similar to the scenario of the original state, indicating the switching of the Néel order back to [100]. Note that such a switching is nonvolatile, because all of the XMLD spectra were measured *in-situ* after removing $E$.

To directly compare the XMLD signals at different states, the values of $\Delta\text{XMLD} = \text{XMLD}_{@638} - \text{XMLD}_{@639}$, defined by the difference between two extreme values (peak or valley) at ~638 and ~639 eV at $L_3$-edge, are listed in Fig. 2B, which are recorded at zero-field after reversibly applying representative electric fields (+4 and −2 kV/cm) to PMN-PT. Remarkably, the $\Delta\text{XMLD}$ value for the $E = +4$ kV/cm case is positive, in contrast to a negative value for the $E = -2$ kV/cm counterpart. As these two electric fields are cycled for three times, the $\Delta\text{XMLD}$ values are changed between positive and negative with well-defined reproducibility. That means, the UMA of the Mn$_2$Au films can be switched electrically between [100] and [0$\bar{1}$1] reversibly, which is also supported by the planar Hall effect (fig. S7). Furthermore, *E*-dependent ΔXMLD values are summarized in Fig. 2C, where *E* are swept from +4 to –4 kV/cm, and then back to +4 kV/cm as a circle (fig. S8). The most eminent feature is that the data points trace a butterfly loop. This feature unravels that the switching of the UMA between [100] and [0$\bar{1}$1] (90° rotation) in the Mn$_2$Au is mainly ascribed to the ferroelastic strain, because it is well established that the *E*-dependent strain evolution (displacement) of the PMN-PT exhibits a butterfly loop.

Figure 3A depicts a schematic of the Mn$_2$Au/PMN-PT device for the NSOT (*19–21*), where the measurement configuration is included. For this experiment, after setting the gating *E*-field to zero, five successive writing current pulses were applied along the [100] channel (write 1) and then along its orthogonal direction, [0$\bar{1}$1] channel (write 2). A small reading current was applied after each writing current pulse, while the transverse resistance was recorded. Figure 3B illustrates the transverse resistance variation (Δ$R_{xy}$) as a function of the number of current pulses for different *E* acting on the PMN-PT. Figure panels (i), (ii), (iii), and (iv) correspond to the original state as well as *E* = +4, –2, and +4 kV/cm scenarios, respectively. For the original state, the five current pulses along the easy-axis [100] set the AFM moments to hard-axis [0$\bar{1}$1], enhancing Δ$R_{xy}$ gradually and uniformly, which is quite characteristic for the current-driven AFM moments switching with a multi-domain process (*19,20,23*). The variation tendency of Δ$R_{xy}$ confirms that the NSOT switching indeed occurs, but is reluctant in this case. Differently, once a single current pulse is applied along [0$\bar{1}$1], Δ$R_{xy}$ shows a sudden drop to a lower value and gets almost saturated, and the following four pulses cause subtle or even negligible Δ$R_{xy}$. This asymmetric character unravels that the AFM moments tend to switch back to the easy-axis [100] with only one current pulse.

The electric field switching of the UMA has a profound influence on the NSOT. After the switching of the Néel order from [100] to [0$\bar{1}$1] via $E$ = +4 kV/cm, $\Delta R_{xy}$ keeps almost constancy with the current pulses along the [100] channel (write 1), because the AFM moments have already been aligned along [0$\bar{1}$1] by the electric field and seldom left along [100]. In contrast to the step-by-step variation of $\Delta R_{xy}$ caused by five successive current pulses along the easy-axis [0$\bar{1}$1] (write 2), $\Delta R_{xy}$ jumps up abruptly and becomes almost saturated with only one current pulse along the [100] channel. Then the following four pulses just affect $\Delta R_{xy}$ slightly. These features are opposite to those of the original state, supporting that the easy-axis changes to [0$\bar{1}$1] induced by the $E$ gating and resultant convenience of the switching from [100] to [0$\bar{1}$1]. With $E$ = –2 kV/cm, current pulses dependent $\Delta R_{xy}$ exhibits a similar tendency as the original case, disclosing that the easy-axis returns to [100], coinciding with the XMLD results in Fig. 2. As $E$ is cycled back to +4 kV/cm again, the NSOT feature behaves analogically as the previous $E$ = +4 kV/cm counterpart, revealing that $E$-induced UMA switching is reversible, reproducible, and reliable. The asymmetric NSOT switching can be considered as a ratchet-like behavior (*24,25*), where the UMA and the same current pulses in the two channels for the NSOT serve as the asymmetric potential and unbiased external input, respectively. As illustrated in Fig. 3C, when the magnetic easy-axis aligns along the [100] direction, the NSOT-driven AFM moments rotation from [0$\bar{1}$1] to [100] is comparatively easier than the opposite rotation. In this case, the [100] easy-axis is the preferential axis of the AFM ratchet. When the easy-axis is set to [0$\bar{1}$1] by the gated field, it is easier for the NSOT switching from [100] to [0$\bar{1}$1]. Then the [0$\bar{1}$1] direction turns out to be the preferential axis of the AFM ratchet.

Figure 4A and 4B compare the current density dependence of $\Delta R_{xy}$ when the current is applied in the [100] (write 1) and [0$\bar{1}$1] (write 2) channels, respectively, for the original state and the gated scenarios with $E$ = +4 and −2 kV/cm. The gated electric field and concomitant ferroelastic strain, strongly affect threshold current density $J_{th}$, at which $\Delta R_{xy}$ starts to jump. In Fig. 4A, where the writing current is applied along the [100] channel, the abrupt change of $\Delta R_{xy}$ occurs at a lower $J_{th}$ of $6.5 \times 10^6$ A cm$^{-2}$ for $E$ = +4 kV/cm compared to those for $E$ = −2 kV/cm and the original state ($J_{th} \approx 8.5 \times 10^6$ A cm$^{-2}$). This roughly 30% change of $\Delta R_{xy}$ is attributed to the $E$-switching of the UMA. When the current is applied in the [0$\bar{1}$1] channel [Fig. 4B], the situation reverses completely: higher $J_{th}$ of ~8.5 $\times 10^6$ A cm$^{-2}$ for the $E$ = +4 kV/cm case, lower $J_{th}$ of ~6.5 $\times 10^6$ A cm$^{-2}$ for the other two states. These contrasting tendencies unambiguously affirm that the easy-axis is aligned to [0$\bar{1}$1] by $E$ = +4 kV/cm, which is beneficial for the NSOT switching from [100] to [0$\bar{1}$1], which can be reversed by $E$ of −2 kV/cm.

We performed numerical simulations of $E$-dependent $J_{th}$ for a single-domain Mn$_2$Au. In Fig. 4C and 4D, the numerically-determined terminal direction of the Néel order is plotted as a function of the writing current density and $E$ for the current being applied in the [100] and [0$\bar{1}$1] channels, respectively. Note that the boundary separating different terminal states and $J_{th}$ change dramatically at $E$ = −2 kV/cm, which agrees well with our experimental observation. A detailed inspection shows that a higher (lower) $J_{th}$ in the [100] ([0$\bar{1}$1]) channel is needed for the NSOT switching from [100] to [0$\bar{1}$1] (from [100] to [0$\bar{1}$1]) as $E$ is swept down to around −2 kV/cm in Fig. 4C (Fig. 4D), corresponding to the experimental case in Fig. 4A (Fig. 4B). Even though a single-domain simulation cannot reproduce the step-like pattern appearing in Fig. 3, it does reveal the essential feature of the strain-induced UMA averaged over all domains. Despite that the absolute value of $J_{th}$ obtained in a single-domain model (~$10^8$ A cm$^{-2}$) cannot be directly converted into the $J_{th}$ of the multi-domain switching (~$10^7$ A cm$^{-2}$) (*27,28*), the features of $J_{th}$ change with *E* sweeping is consistent with the experimental counterpart. We conclude by noting that the electric field switching of the uniaxial magnetic anisotropy is realized in antiferromagnetic Mn$_2$Au films at room temperature, thus promising electric field-tunable shift registor (*24,25*), magnetic memory (*26*), spin current on-off (*29,30*) and spin logic devices (*8*) with low power consumption.

ACKNOWLEDGMENTS

We are grateful to the fruitful discussions with J. H. Han and Dr. D. Z. Hou. C.S. acknowledges the support of Beijing Innovation Center for Future Chip (ICFC) and Young Chang Jiang Scholars Program. The XMLD measurements were carried out at Beamline BL08U1A of SSRF. This work was supported by the National Key R&D Program of China (Grant Nos. 2017YFB0405704) and the National Natural Science Foundation of China (Grant Nos. 51871130, 51571128 and 51671110).

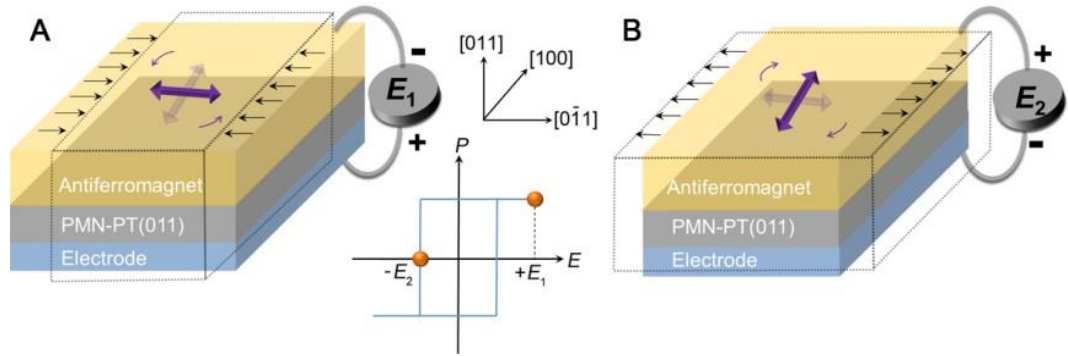

**Fig. 1 Schematic of ferroelastic strain switching of UMA driven by electric fields in antiferromagnet/PMN-PT(011) structure.** AFM moments are switched towards the compressive direction of the PMN-PT substrate, [0$\bar{1}$1] and [100] axes in (**A**) and (**B**), respectively. The axes denote the crystalline directions of (011)-oriented PMN-PT substrate. The inset shows a cartoon for the *P–E* loop.

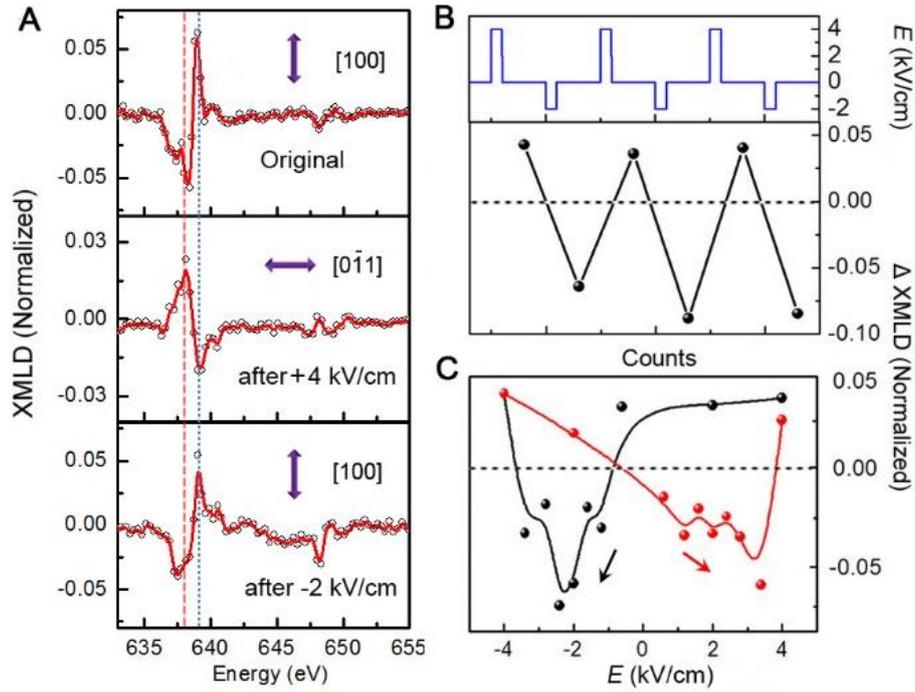

**Fig. 2 *E*-dependent Mn *L*–edge XMLD signals of $Mn_2Au$.** (**A**) Mn *L*–edge XMLD spectra of $Mn_2Au$ at the original state and after applying *E* of +4 and –2 kV/cm to the PMN-PT(011). X-ray is vertically incident to the $Mn_2Au$ film. Corresponding Néel order is marked in the inset. (**B**) ΔXMLD values after reversibly applying representative *E* of +4 and –2 kV/cm to the PMN-PT(011). (**C**) Summarized ΔXMLD values as a function of *E*, which are swept from +4 to –4 kV/cm, and then back to +4 kV/cm as a circle. The solid line is just a guide to the eyes.

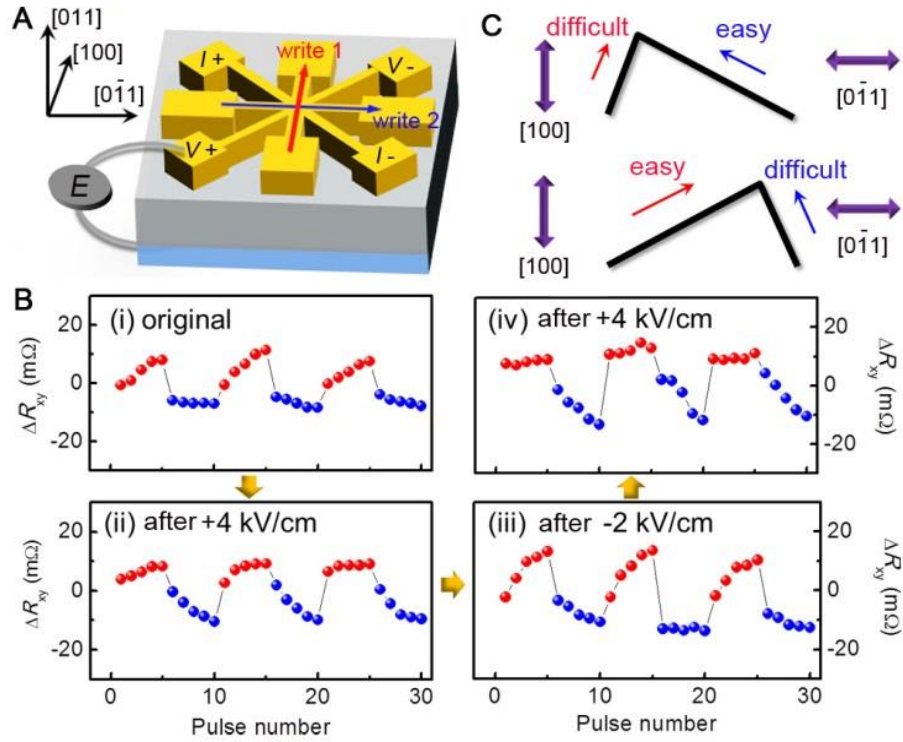

**Fig. 3 Ratchet-like NSOT in Mn$_2$Au with UMA**. (**A**) Measurement configurations for the NSOT. Writing current pulses along the [100] and [0$\bar{1}$1] channels are denoted as write 1 and 2, respectively. Current *I* and voltage *V* show the readout scheme. (**B**) $\Delta R_{xy}$ of the Mn$_2$Au device as a function of the number of current pulses after applying different *E*. The original state as well as *E* = +4, –2, and +4 kV/cm scenarios are shown in panel (i), (ii), (iii), and (iv), respectively, in order of measurement sequence. The colors of symbols for the $\Delta R_{xy}$ data correspond to the writing current channels. (**C**) Schematics of the ratchet-like behavior based on the asymmetric NSOT switching when the magnetic easy axes are aligned long the [100] (top panel) or [0$\bar{1}$1] (bottom panel) direction.

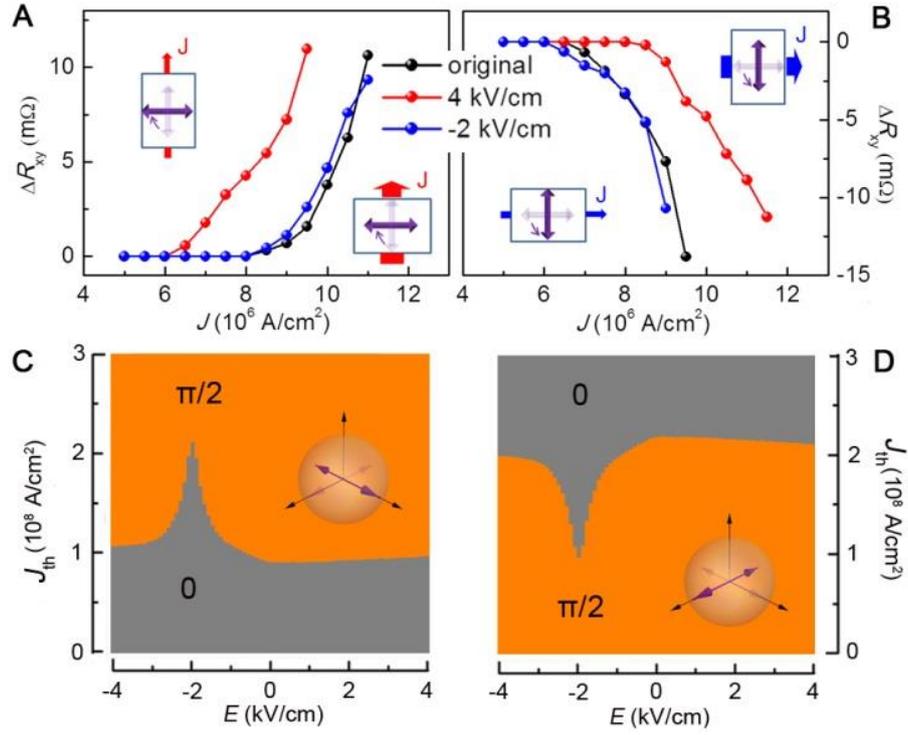

**Fig. 4 $E$-control of $J_{th}$ for NSOT in Mn$_2$Au**. Summarized $\Delta R_{xy}$ as a function of current density when the current is applied in the [100] (**A**) and [0$\bar{1}$1] channels (**B**). $\Delta R_{xy}$ was recorded for the original state or after applying representative $E$ (+4 and –2 kV/cm) on the PMN-PT. Simulations of $E$-dependent $J_{th}$ for Mn$_2$Au single-domain switching from [100] towards [0$\bar{1}$1] (**C**) and its opposite direction (**D**). The applied current density lower and higher than $J_{th}$ drives no switching and 90° switching ($\pi/2$) of the AFM domain, respectively.

13